\documentclass{aastex}
\usepackage{emulateapj5}
\usepackage{apjfonts}
\usepackage{psfig}

\slugcomment{Accepted by the Astrophysical Journal Letters}
\shorttitle{Black Holes: Spins and Evolution}
\shortauthors{Wang, Chen, Ho \& McLure}

\def\kms{\ifmmode {\rm km~ s^{-1}} \else {\rm km~s^{-1}}\ \fi}
\def\mbh{M_{\bullet}}
\def\mbhc{M_{\bullet}^{\rm c}}
\def\mgii{\ifmmode Mg {\sc ii} \else Mg {\sc ii}\ \fi}
\def\rhobh{\rho_{\bullet}}
\def\rmd{{\rm d}}

\def\sunm{M_{\odot}}

\def\lax{{$\mathrel{\hbox{\rlap{\hbox{\lower4pt\hbox{$\sim$}}}\hbox{$<$}}}$}}
\def\gax{{$\mathrel{\hbox{\rlap{\hbox{\lower4pt\hbox{$\sim$}}}\hbox{$>$}}}$}}

\begin{document}

%\title{Black Holes in Quasars from Sloan Digital Sky Survey: Spin and Its Evolution}
\title{Evidence for Rapidly Spinning Black Holes in Quasars}

\author{Jian-Min Wang\altaffilmark{1}, Yan-Mei Chen\altaffilmark{1,2}, Luis C. Ho\altaffilmark{3} 
and Ross J. McLure\altaffilmark{4} }

\altaffiltext{1}{Key Laboratory for Particle Astrophysics, Institute of High Energy Physics, 
                 Chinese Academy of Sciences, 19B Yuquan Road, Beijing 100049, China}

\altaffiltext{2}{Graduate School, Chinese Academy of Science, 19A Yuquan Road, Beijing 100049, China}

\altaffiltext{3}{The Observatories of the Carnegie Institution of Washington, 813 Santa Barbara Street, Pasadena, CA 91101, USA}

\altaffiltext{4}{Institute for Astronomy, University of Edinburgh, Royal Observatory, Edinburgh EH9 
                 3HJ, UK}

\begin{abstract}
It has long been believed that accretion onto supermassive black holes powers 
quasars, but there has been relatively few observational constraints on the 
spins of the black holes.  We address this problem by estimating the average
radiative efficiencies of a large sample of quasars selected from the 
Sloan Digital Sky Survey, by combining their luminosity function and their 
black hole mass function.  Over the redshift interval $0.4<z<2.1$, we find that
quasars have average radiative efficiencies of $\sim 30\% - 35\%$, strongly 
suggesting that their black holes are rotating very fast, with specific 
angular momentum $a \approx 1$, which stays roughly constant with redshift.  
The average  radiative efficiency could be reduced by a factor of $\sim$2, 
depending on the adopted zeropoint for the black hole mass scale.
The inferred large spins and their lack of significant evolution are in 
agreement with the predictions of recent semi-analytical models of 
hierarchical galaxy formation if black holes gain most of their mass through 
accretion.
\end{abstract}
 
\section{Introduction}
Supermassive black holes are generally believed to be the power sources in 
quasars and other active galactic nuclei (Rees 1984), and in recent years 
there has been tremendous progress not only in measuring their masses but also
in linking them to the global properties of their host galaxies (see reviews 
in Ho 2004).  Apart from the mass, the other fundamental property of 
astrophysical black holes is the spin. However, to date there have been 
relatively few observational constraints.  A handful of Seyfert galaxies, 
the most notable being MCG~$-$6-30-15 (Wilms et al. 2001; Fabian et al. 2002), 
show a relativistically broadened, highly redshifted iron K$\alpha$ line that 
can be most plausibly be interpreted as arising from a compact region around 
a rapidly rotating black black hole.  
%The broad iron  K$\alpha$ line, on the other hand, for reasons that are not 
%yet fully clear, is now know to be a relatively rare feature in Seyfert 
%galaxies (e.g., Bianchi et al. 2004).   
The quasi-periodic variability detected in Sgr A*, both in the near-infrared 
and in the X-rays, can also be interpreted as evidence for a large spin for 
the Galactic Center black hole (Genzel et al. 2003; Ashenbach et al. 2004). 
Spectral fitting of the broad-band X-ray spectrum of active galaxies 
has achieved particular success by invoking ionized reflection disk models 
with inner disk radii sufficient compact to suggest maximally rotating black 
holes (Crummy et al. 2006).  Lastly, mild evidence for rotating black holes
has come from integral constraints derived for global populations of active 
galaxies.  Yu \& Tremaine (2002), applying So\l tan's (1982) argument to a 
sample of $z\approx 0-5$ quasars, concluded that their high average radiative 
efficiency ($\bar{\eta}$ \gax\ 0.1) implies that their black holes
are spinning.  Elvis et al. (2002) applied a similar calculation to the cosmic 
X-ray background and concluded that $\bar{\eta}$ \gax\ 0.15.

Theoretical considerations do not provide a clear prediction of the 
observational expectation.  While gas accretion inevitably increases the 
spin of the black hole, as do mergers of comparable-mass black holes under 
most circumstances (Volonteri et al. 2005), minor mergers tend to have the 
opposite effect (Hughes \& Blandford 2003; Gammie et al. 2004; Volonteri et 
al. 2005).  Thus, the spin of the black hole of any given galaxy at any 
particular time depends on its specific merger history up to that point.

In this Letter, we attempt to constrain the spins of supermassive black holes 
by estimating the average radiative efficiency of a large sample of quasars 
with redshifts $z=0.4-2.1$.  The underlying assumption of our method is that 
black holes attain most of their mass through accretion.  We find that quasars
radiate with a high efficiency ($\bar{\eta} \approx 0.3-0.35$), from 
which we infer that their black holes are rapidly rotating.  Throughout, our 
calculations assume the following cosmological parameters: 
$H_0=70~{\rm Mpc^{-1}~km~s^{-1}}$, $\Omega_{\rm M}=0.3$, and 
$\Omega_{\Lambda}=0.7$.

\begin{figure*}
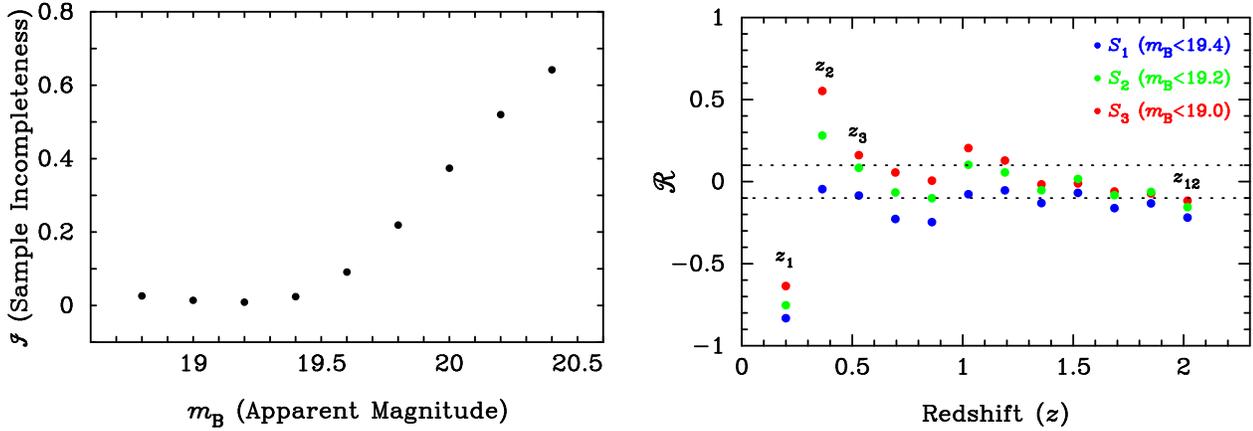

{\vbox
{\hbox{
\psfig{figure=fig1a.ps,angle=270,width=8cm}
\hskip+0.2in
\psfig{figure=fig1b.ps,angle=270,width=8cm}
}
}}
\vglue 0.15cm
\caption{\footnotesize 
({\it Left}) The global incompleteness of the sample as a function of apparent 
magnitude.  Note that the sample becomes increasingly incomplete for 
$m_{\rm B}$ \gax $19.5 - 19.6$ mag.  ({\it Right}) Test of the completeness of 
each of the three samples, for different redshift bins. The bins $z_1$ and 
$z_2$ poorly match the SDSS quasar luminosity function of Richards et al. 
(2006). This can be attributed to the inhomogeneity of the McLure \& Dunlop 
(2004) sample in the redshift interval $z=0.117-0.4$; these two bins are 
excluded from the analysis.  The other redshift bins have a completeness level 
of at least 98\%.  
}
\label{fig1}
\end{figure*}

\section{Accretion-growth Equation and Radiative Efficiency }
If quasar light derives from accretion of matter onto a black hole, then the 
radiative efficiency is $\eta\approx \Delta \epsilon/\Delta \rhobh c^2$, where 
the black hole mass density increase $\Delta \rhobh$ in a redshift interval 
$\Delta z$ at $z$ results in an increase of the radiative energy density 
$\Delta \epsilon$.  In practice, $\Delta \epsilon$ can be derived from the 
quasar luminosity function and $\Delta \rhobh$ can be obtained from the mass 
distribution function.  Thus, we can estimate the radiative efficiency at any 
redshift, and hence place a strong constraint on the average spin of black 
holes, since $\eta$ varies as a function of spin ($\sim 0.06$ and 0.42 for a 
non-rotating and a maximally rotating black hole, respectively).  If black 
hole masses are known for a quasar sample, we can define their mass 
distribution function as 
\begin{equation}
\Phi(\mbh,z)=\frac{\rmd^2 N}{\rmd\mbh\rmd V}, 
\end{equation}
where $\mbh$ is the black hole mass, $\rmd N$ is the number of quasars within the comoving volume 
element $\rmd V$ and mass interval $\rmd \mbh$.  Then the integrated mass density of black holes 
with $\mbh\ge \mbhc$ at a redshift $z$ is given by 
\begin{equation}
\rhobh(z)=\int_z^{\infty}\rmd z\int_{\mbhc}^{\infty}\mbh \Phi(\mbh,z)\rmd\mbh,
\end{equation}
where $\mbhc$ is a lower limit set by the flux limit of the survey.

Accretion of matter onto a black hole generates radiation and increases 
the mass of the hole.  The relationship between the radiated energy and the 
accumulated mass density in black holes can be expressed by the 
accretion-growth equation as
\begin{equation}
\rhobh(z) = \int_z^{\infty}\frac{\rmd t}{\rmd z} \rmd z\int_{L_{\rm min}(z)}^{\infty}
            \frac{(1-\eta)}{\eta} \frac{L_{\rm bol}}{c^2}\Psi(L,z)\rmd L,
\end{equation} 
where $L_{\rm min}(z)$ is the minimum luminosity of the survey at redshift 
$z$, $c$ is the speed of light, $L_{\rm bol}$ is the bolometric luminosity, 
$\Psi(L,z)$ is the luminosity function of the quasar sample, $L$ is the 
specific luminosity, and the radiative efficiency $\eta$ varies with redshift. 
Equation (3) involves a relation between the bolometric and specific 
luminosities.  We convert the $B$-band luminosity $L_{\rm B}$ into the 
bolometric luminosity via $L_{\rm bol}=C_{\rm B} L_{\rm B}$, where 
$C_{\rm B}\approx 6-7$ is the $B$-band bolometric correction factor for 
quasars brighter than $L_{\rm bol}\approx 10^{11.5}\,L_{\odot}$ (Marconi et 
al. 2004); we adopt $C_{B}=6.5$.  Since the spectral energy distributions of 
quasars show little evidence for redshift evolution (Shemmer et al. 2005; 
Strateva et al. 2005; Steffen et al. 2006), we assume that $C_{\rm B}$ does 
not vary with redshift.  With $\Phi(\mbh,z)$ and $\Psi(L,z)$, the average 
radiative efficiency $\bar{\eta}(z)$ at each redshift bin follows from the 
differential version of Equation (3): 
\begin{equation}
\bar{\eta}(z)=\frac{\Delta\epsilon}{\Delta \epsilon+\Delta\rhobh c^2},
\end{equation}
where 
\begin{equation}
\Delta \epsilon=\Delta z~\frac{\rmd t}{\rmd z}\int_{L_{\rm min}(z)}^{\infty}L_{\rm bol}\Psi(L,z)\rmd L
\end{equation}
and 
\begin{equation}
\Delta\rhobh=\Delta z\int_{\mbhc}^{\infty}\mbh\Phi(\mbh,z)\rmd \mbh.
\end{equation}
Assuming that the innermost stable circular orbit is the inner radius of the 
accretion disk (Bardeen et al. 1972), the inferred radiative efficiency yields 
an estimate of the black hole spin.  

Equation (4) constrains the radiative efficiency at {\em any} redshift,
provided the black hole mass function is known. It should be stressed that 
both sides of Equation (3) only include the actively black holes (i.e. 
quasars), and thus the radiative efficiency from Equation (4) does not rely on 
the lifetime of quasars. This method is also independent of obscured sources, 
which is an important complicating factor in estimation of the radiative 
efficiency using So\l tan's method (Elvis et al. 2002; Yu \& Tremaine 2002).

\section{Application to SDSS Quasars}
\subsection{Estimation of black hole masses}

The large database provided by the Sloan Digital Sky Survey (SDSS; York et al. 
2000) affords us an excellent opportunity to examine this problem.  
Reverberation mapping of local active galaxies has resulted in empirical 
scaling relations based on quasar luminosity and broad emission line width 
(Kaspi et al. 2000; McLure \& Jarvis 2002; Vestergaard 2002) that enable 
``virial'' black hole masses to be obtained, and hence the distribution 
function for the black hole masses of a sample of quasars can be estimated 
independently from their luminosity function.   Because the profile of the 
C~{\sc iv} line is complex and may be strongly affected by outflows (Baskin 
\& Laor 2005), we only consider objects with broad H$\beta$ and \mgii\ lines 
detected in SDSS.  This limits the maximum redshift of the present sample to 
$z\le 2.1$.  For quasars with $z\le 0.7$, we obtain the virial black hole 
masses using the FWHM of the H$\beta$ line ($V_{\rm H\beta}$) following the 
empirical relation\footnote{This relation for H$\beta$ is based on the 
original work of Kaspi et al. (2000), which has since been recalibrated 
(Onken et al. 2004; Kaspi et al. 2005). The new calibration increases the 
zeropoint of the mass scale by roughly a factor of 2.  However, the new 
zeropoint has not been established with great statistical certainty (Nelson et 
al. 2004; Greene \& Ho 2006), and for the current application we will retain 
the original zeropoint of Kaspi et al.  (2000), on which the masses derived by 
McLure \& Dunlop (2004) are based.}
\begin{equation}
\mbh=4.7\times10^6\left(\frac{L_{5100}}{10^{37}{\rm W}}\right)^{0.61}
\left(\frac{V_{\rm H\beta}}{10^3~{\rm km}~{\rm s}^{-1}}\right)^2 \, \sunm, 
\end{equation}
where $L_{5100}$ is the specific continuum luminosity at 5100 \AA. For higher 
redshift quasars ($0.7<z\le 2.1$), we use the FWHM of Mg~{\sc ii} 
($V_{\rm Mg~II}$) to estimate the black hole mass, using the calibration 
(McLure \& Dunlop 2004) 
\begin{equation}
\mbh=3.2\times10^6\left(\frac{L_{3000}}{10^{37}{\rm W}}\right)^{0.62}
\left(\frac{V_{\rm Mg~II}}{10^3~{\rm km}~{\rm s}^{-1}}\right)^2 \, \sunm, 
\end{equation}
where $L_{3000}$ is the specific continuum luminosity at 3000 \AA. The scatter 
of these relations has been estimated to be $\sim$0.4 dex (McLure \& Dunlop 
2004).

\subsection{Samples}
Our analysis is based on black hole masses calculated by McLure \& Dunlop 
(2004) for 12,698 quasars in the redshift range $0.1\le z\le 2.1$, for which 
good-quality spectra are available from the quasar catalog of the First Data 
Release (DR1) of SDSS (Schneider et al. 2003).  This sample, however, is 
neither complete nor homogeneous.  To evaluate its completeness, we compare it 
to the quasar luminosity function recently determined for the Third Data 
Release (DR3) of SDSS (Richards et al. 2006). After dividing the sample into 
$n_z=12$ redshift bins, we evaluate the two parameters 
\begin{equation}
{\cal R}_i=\frac{N_{\rm bin}^i-N_{\rm LF}^i}{N_{\rm LF}^i};~~~~~
{\cal I}=\sum_{i=1}^{n_z}{\cal R}_i^2, 
\end{equation}
where $N_{\rm bin}^i$ is the number of quasars within the redshift bin $z_i$ 
and $z_i+\Delta z$ and $N_{\rm LF}^i$ is the number of quasars calculated from 
the luminosity function. The parameter ${\cal R}_i$ measures the degree of 
incompleteness of the sample at each redshift bin $z_i$, and ${\cal I}$ 
indicates the global incompleteness of the sample.  By adjusting the apparent 
magnitude $m_{B}$, we can define subsamples with different levels of 
completeness.  As illustrated in Figure 1 ({\it left}), the sample 
incompleteness begins to be noticeable for $m_{\rm B}$ \gax 19.5--19.6 mag.  
Furthermore,  Figure 1 ({\it right}) shows that the number of quasars in the 
first two redshift bins matches poorly with the predictions based on 
the luminosity function of Richards et al. (2006). We thus restrict our 
attention to the redshift range $0.4\le z\le 2.1$, and consider only three 
subsamples, $S_1$, $S_2$ and $S_3$, which correspond to apparent magnitude 
limits $m_{\rm B}<19.4$, 19.2, and 19.0 mag, respectively. The completeness 
level of these subsamples is \gax 98\%.

\subsection{Results}
We show the differential black hole mass density ($\rmd \rhobh/\rmd z$) as a 
function of redshift in Figure 2{\it a}. We find that the black hole density 
is very sensitive to the limit magnitudes of the samples.  The quasar 
luminosity function from SDSS DR3, as given by Richards et al. (2006), is 
\begin{equation}
\Psi=\Psi_*10^{A_1\left[M_i-\left(M_i^*+B_1\xi+B_2\xi^2+B_3\xi^3\right)\right]},
\end{equation}
where $M_i$ is $i-$band magnitude, 
$\xi=\log\left[(1+z)/(1+z_{\rm ref})\right]$, $A_1=0.84$, $B_1=1.43$, 
$B_2=36.63$, $B_3=34.39$, $M_i=-26$, $z_{\rm ref}=2.45$ and 
$\Psi_*=10^{-5.7}$.  Inserting the luminosity function and the mass 
distribution function into Eq. (4), we immediately arrive at the radiative 
efficiency plotted in Figure 2{\it b}. Independent of the chosen subsample, we 
find that quasars radiate at a high efficiency, with $\eta \approx 0.3-0.35$, 
roughly independent on redshift from $z\approx 0.4$ to  $z\approx 2$. 
The average radiative efficiency we obtain is significantly higher than that 
corresponding to the maximum spin ($a\approx 0.9$) achieved by a 
magnetohydrodynamic thick disk (Gammie et al. 2004), $\eta\approx 0.2$, but is 
consistent with 
the efficiency of Thorne's (1974) limit of $a=0.998$ ($\eta=0.32$), as well 
as that of extreme Kerr rotation $a=1$ ($\eta=0.42$).  A similar conclusion 
has been reached by other studies, based on models of hierarchical galaxy 
formation (Volonteri et al. 2005) and considerations of the cosmic X-ray 
background radiation (Elvis et al. 2002).  We should note that Elvis et al.  
(2002) only gave a lower limit on the efficiency ($\eta>0.15$), and no 
detailed information at any given redshift. The observed high value and 
constancy of $\eta$ suggests the spin angular momentum of most or all black 
holes in quasars have saturated at their maximum value and undergo no 
evolution from $z\approx 2$ to $z\approx 0.4$. Our results are consistent with
the conclusions of Volonteri et al. (2005).

\figurenum{2}
\centerline{\psfig{figure=fig2.ps,angle=270,width=7.5cm}}
\vglue 0.2cm
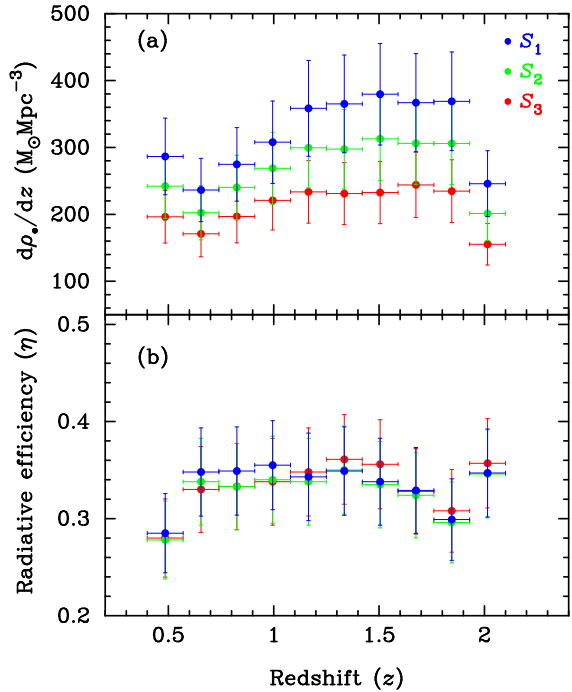
\figcaption{\footnotesize The black hole mass density ({\it a}) and the
radiative efficiency ({\it b}) as a function of redshift for three quasar
samples ($S_1$, $S_2$ and $S_3$) from the SDSS DR1. The error bars
on $\rmd \rho_{\bullet}/\rmd z$ and $\eta$ are dominated by the uncertainty  
in the black hole masses.}
\label{fig2}

\section{Discussion and Summary}

We have estimated the average radiative efficiency, and hence the spin, of 
supermassive black holes by combining the luminosity and black hole mass 
function of a large sample of SDSS quasars selected over the redshift interval 
$0.4<z<2.1$.  With find that the average radiative efficiency is very high, 
$\bar{\eta} \approx 0.3-0.35$, which implies that the black holes are
rotating very fast, with $a\approx 1$.  No noticeable evolution is seen over 
this range of redshifts; it would be very interesting to extend this study 
to higher redshifts (\gax 2) in order to establish the epoch over which the 
black hole spins were imprinted.  We note that our conclusions are only 
based on the accretion-growth equation, which makes no reference to whether 
the black mass was gained principally through accretion or mergers.  An 
advantage of the present approach is that the result does not depend on the 
lifetime of quasars. 

The large spins deduced for the black holes in quasars may arise quite 
naturally as a consequence of major (nearly equal-mass) galaxy mergers.  
Massive, gas-rich mergers account not only for most of the star formation in 
the Universe at $z\approx 2-3$ (e.g., Conselice et al. 2003), but they are 
probably also responsible for triggering major episodes of quasar activity 
(Di~Matteo et al.  2005).  When galaxies merge, so, too, do their black holes 
(if they exist in the parent galaxies), at least in principle.  The spin 
angular momentum of black holes newly born from mergers is expected to be 
high, with $a>0.8$, its exact value depending on the orbit of the original 
binary (Gammie et al.  2004).  Subsequent accretion at an Eddington-limited 
rate will further increase the spin on a timescale shorter than the Salpeter 
time ($\tau=0.45$ Gyr) (Volonteri et al. 2005). 
% XX you originally referenced Thorne 1974 here.  What is this in reference to?
% is it the Salpeter time (which is trivial, and required no reference) or
% is it for the statement that accretion further increases the spin on a short
% timescale?  If the latter, should not the proper reference be Volenteri?
The newborn holes can thus rapidly evolve into Kerr holes, consistent with 
our results.

The high radiative efficiency prolongs the lifetime of a quasar's 
accretion. Following Shapiro (2005), the mass of an accreting black hole 
at time $t$ with an initial mass $M_0$ is given by
\begin{equation}
\mbh(t)=M_0\exp\left[\frac{\dot{m}(1-\eta)}{\eta}\frac{t}{\tau}\right],
\end{equation}  
where $\dot{m}=\dot{M}_{\rm acc}c^2/L_{\rm Edd}$, $\dot{M}_{\rm acc}$ is the 
accretion rate, and $L_{\rm Edd}$ is the Eddington luminosity. The anticipated 
lifetime of an $e-$fold accretion-growth is then $t_{\rm QSO}\approx\eta 
\tau/\dot{m}(1-\eta)\approx (0.4-0.7)\tau \approx 0.2-0.3$ Gyr 
%
%XX you originally had 0.5-0.7; I think it should be 0.4-0.7
%
for $\eta=0.3-0.4$ if the hole is accreting at the Eddington rate 
($L_{\rm Edd}/c^2$). This lifetime lies within the range of values currently 
estimated for quasars (Martini 2004), but is uncomfortably long for the 
highest-redshift quasars known ($z$ \gax 6.4), whose large masses 
($\mbh \approx 10^9\, \sunm$), if grown from much smaller seeds, would 
have required much more rapid growth rates (Shapiro 2005).

One major caveat affects the actual numerical value of $\bar{\eta}$.  We noted 
in \S 3.1 that our black hole mass scale is based on the original zeropoint of 
Kaspi et al. (2000), whose accuracy is currently still a subject of debate.
If the zeropoint in fact should be increased by a factor of $\sim$2, as 
suggested by some recent studies, then the radiative efficiency would 
{\it decrease}\ by the same factor, to $\bar{\eta} \approx 0.2$.   This lower 
value of $\bar{\eta}$ would help alleviate the conflict with the growth rate
of the highest-redshift quasars (Shapiro 2005), would be consistent with the
theoretical expectation of magnetohydrodynamic thick disks (Gammie et al. 
2004), and would bring it in better agreement with
the independent estimates of radiative efficiencies derived from So\l tan-type
arguments.  The analysis of optically selected quasars by Yu \& Tremaine 
(2002) obtains $\bar{\eta}$ \gax 0.1 assuming a bolometric correction 
of $C_B=11.8$.  If $C_B=6.5$ had been adopted, as suggested by Marconi et al. 
(2004), Yu \& Tremaine's value of $\bar{\eta}$ would be lower by a factor of 
$\sim 2$.  On the other hand, this type of estimate is seriously affected 
by uncertainties in the contribution from obscured active galaxies, which is 
currently poorly known (Martinez-Sansigre et al. 2005).  Estimates of the 
average radiative efficiency using an energy density based on the cosmic 
X-ray background, which is less affected by obscuration, yield values of 
$\bar{\eta}$ (\gax 0.15; Elvis et al. 2002) that are more consistent with our 
results.

\acknowledgements{We thank the referee for helpful comments on 
the manuscript.  J. M. W. acknowledges support from a Grant for Distinguished 
Young Scientist from NSFC (NSFC-10325313, NSFC-10233030, NSFC-10521001), L. C. H. from 
NASA and the Carnegie Institution of Washington, and R. J. M from the Royal 
Society.}

\end{document}